\documentclass[a4paper,12pt]{article}
\usepackage{latexsym}
\makeatletter

\newcommand{\nn}{\nonumber\\}\newcommand{\p}[1]{(\ref{#1})}

\def\PRL #1 #2 #3{{\em Phys. Rev. Lett. \/} {\bf#1} (#2) #3}
\def\NPB #1 #2 #3{{\em Nucl. Phys. \/} {\bf B#1} (#2) #3}
\def\NPBFS #1 #2 #3 #4{{\em Nucl. Phys. \/} {\bf B#2} [FS#1] (#3) #4}
\def\CMP #1 #2 #3{{\em Commun. Math. Phys. \/} {\bf #1} (#2) #3}
\def\PRD #1 #2 #3{{\em Phys. Rev. \/} {\bf D#1} (#2) #3}
\def\PLA #1 #2 #3{{\em Phys. Lett. \/} {\bf #1A} (#2) #3}
\def\PLB #1 #2 #3{{\em Phys. Lett. \/} {\bf B#1} (#2) #3}
\def\JMP #1 #2 #3{{\em J. Math. Phys. \/} {\bf #1} (#2) #3}
\def\PTP #1 #2 #3{{\em Prog. Theor. Phys. \/} {\bf #1} (#2) #3}
\def\SPTP #1 #2 #3{{\em Suppl. Prog. Theor. Phys. \/} {\bf #1} (#2) #3}
\def\AoP #1 #2 #3{{\em Ann. of Phys. \/} {\bf #1} (#2) #3}
\def\PNAS #1 #2 #3{{\em Proc. Natl. Acad. Sci. USA} {\bf #1} (#2) #3}
\def\RMP #1 #2 #3{{\em Rev. Mod. Phys. \/} {\bf #1} (#2) #3}
\def\PR #1 #2 #3{{\em Phys. Reports \/} {\bf #1} (#2) #3}
\def\AoM #1 #2 #3{{\em Ann. of Math. \/} {\bf #1} (#2) #3}
\def\UMN #1 #2 #3{{\em Usp. Mat. Nauk \/} {\bf #1} (#2) #3}
\def\FAP #1 #2 #3{{\em Funkt. Anal. Prilozheniya \/} {\bf #1} (#2) #3}
\def\FAaIA #1 #2 #3{{\em Functional Analysis and Its Application \/} {\bf
#1} (#2) #3}
\def\BAMS #1 #2 #3{{\em Bull. Am. Math. Soc. \/} {\bf #1} (#2)
#3} \def\TAMS #1 #2 #3{{\em Trans. Am. Math. Soc. \/} {\bf #1}
(#2) #3}
\def\InvM #1 #2 #3{{\em Invent. Math. \/} {\bf #1} (#2) #3}
\def\LMP #1 #2 #3{{\em Letters in Math. Phys. \/} {\bf #1} (#2) #3}
\def\IJMPA #1 #2 #3{{\em Int. J. Mod. Phys. \/} {\bf A#1} (#2) #3}
\def\AdM #1 #2 #3{{\em Advances in Math. \/} {\bf #1} (#2) #3}
\def\RMaP #1 #2 #3{{\em Reports on Math. Phys. \/} {\bf #1} (#2) #3}
\def\IJM #1 #2 #3{{\em Ill. J. Math. \/} {\bf #1} (#2) #3}
\def\APP #1 #2 #3{{\em Acta Phys. Polon. \/} {\bf #1} (#2) #3}
\def\TMP #1 #2 #3{{\em Theor. Mat. Phys. \/} {\bf #1} (#2) #3}
\def\JPA #1 #2 #3{{\em J. Physics \/} {\bf A#1} (#2) #3}
\def\JSM #1 #2 #3{{\em J. Soviet Math. \/} {\bf #1} (#2) #3}
\def\MPLA #1 #2 #3{{\em Mod. Phys. Lett. \/} {\bf A #1} (#2) #3}
\def\JETP #1 #2 #3{{\em Sov. Phys. JETP \/} {\bf #1} (#2) #3}
\def\JETPL #1 #2 #3{{\em  Sov. Phys. JETP Lett. \/} {\bf #1} (#2) #3}
\def\PHSA #1 #2 #3{{\em Physica} {\bf A#1} (#2) #3}
\def\CQG #1 #2 #3{{\em Class. Quantum Grav. \/} {\bf #1} (#2) #3}
\def\SJNP #1 #2 #3{{\em Sov. J. Nucl. Phys. (Yadern.Fiz.) \/} {\bf #1} (#2) #3}

\begin{document}

\title{
\begin{flushright}
{\small {DFPD 01/TH/22\\
\vspace{-15pt}
 hep-th/0106212
} }
\end{flushright}
~\\
{Coincident (Super)--Dp--Branes of Codimension One}
 }

\bigskip
\author{
Dmitri Sorokin
~\\
~\\
{\it Institute for Theoretical Physics, NSC KIPT,
Kharkov, Ukraine}\\
{\it and}\\
{\it Universit\`a Degli Studi Di Padova,
Dipartimento Di Fisica ``Galileo Galilei''}\\
{\it ed INFN, Sezione Di Padova Via F. Marzolo, 8, 35131 Padova,
Italia}\\
}
\date{}

\maketitle

\abstract{ We consider properties of a covariant worldvolume
action for a system of N coincident Dp--branes in D=(p+2)
dimensional space--time (so called codimension one branes). In the
case of N coincident D0--branes in D=2 we then find a
generalization of this action to a model which includes fermionic
degrees of freedom and is invariant under target--space
supersymmetry and worldline kappa--symmetry. We find that the type
IIA $D=2$ superalgebra generating the supersymmetry
transformations of the ND0--brane system acquires a non--trivial
``central extension" due to a nonlinear contribution of $U(N)$
adjoint scalar fields. Peculiarities of space--time symmetries of
coincident Dp--branes are discussed.}

\bigskip
\noindent
\renewcommand{\thefootnote}{\arabic{footnote}}
\thispagestyle{empty}
\newpage
\section{Introduction}
Systems of N coincident Dirichlet p--branes play an important role
in String Theory. In particular, their low energy dynamics gives
rise to effective supersymmetric non--Abelian (Born--Infeld--type)
field theories with an internal gauge group U(N). This is why
intensive research have been undertaken to get a detailed
information about the structure and the dynamics of these
multibrane configurations. Such an information can be obtained if
one knows the worldvolume actions which describes the low energy
behaviour of the Dp--branes.

Actions for bosonic systems of N coincident Dp--branes have been
constructed in \cite{Tseytlin:1999dj,Taylor:2000pr,Myers:1999ps}
using T--duality tools and comparison with results known from
Matrix Theory.

These actions are written in a static gauge in which (p+1)
coordinates of the target--space are identified with the
worldvolume coordinates
$\sigma^a$ ($a=0,1,\cdots,p$), and target--space indices
$M,N...$ split into the worldvolume indices $a,b...$ and the indices
$i,j\,...=1,\cdots,D-p-1$ of target--space coordinates transverse to the
D--branes.  Thus the space--time symmetry of the background is
explicitly broken down to a direct product of the worldvolume
symmetry times an internal symmetry acting on the indices
$i,j\,...$ .

The worldvolume action for N coincident Dp--branes in a
D--dimensional bosonic supergravity background has the following
form
\begin{equation}\label{1}
{S}=-T_p \int d^{p+1}\sigma\,{\rm Tr}\,e^{-\phi}\sqrt{-det\left(
P\left[E_{ab}+E_{ai}(Q^{-1}-\delta)^{ij}E_{jb}\right]+
F_{ab}\right)\,det(Q^i{}_j)}+S_{WZ}
\ ,
\end{equation}
where $T_p$ is the brane tension and $\phi$ is the dilaton field.
$F_{ab}=\partial_a A_b-\partial_b A_a+{i\over{2\pi l^2_s}}[A_a,A_b]$ is the field
strength of a worldvolume $U(N)$ gauge field $A_a(\sigma)$. In
addition, the action (\ref{1}) depends on $D-p-1$ scalar fields
$\Phi^i(\sigma)$  and on their covariant derivatives
$D_a\Phi^i=\partial_a\Phi^i+{i\over{2\pi l^2_s}}[A_a,\Phi^i]$
taking values in the
space of the adjoint representation of
$U(N)$ (i.e. they are $N\times N$ Hermitian matrices)\footnote{
In our convention $A_a$ and $\Phi^i$ have the
dimension of length in the string scale $l_s$, in contrast to
\cite{Myers:1999ps} where their dimension is $l^{-1}_s$.}.

$P$ denotes the pull back onto the worldvolume of the matrix
\begin{equation}\label{E}
E_{MN}=G_{MN}+B_{MN}
\end{equation}
composed of the target--space metric $G_{MN}$ and an NS--NS field
$B_{MN}$. The explicit form of the pullback
$P$ of $E_{MN}$
\p{E} is
\begin{equation}\label{PE}
P[E]_{ab}=E_{ab}+\,E_{ai}\,D_{b}\Phi^i+\,D_{a}\Phi^i\,E_{ib}+
\,D_a\Phi^iD_b\Phi^j\,E_{ij} .
\end{equation}
The matrix $Q^i_{~j}$ in \p{1} has the form
\begin{equation}\label{2}
Q^i{}_j\equiv\delta^i{}_j+{i\over{2\pi l^2_s}}\,[\Phi^i,\Phi^k]\,E_{kj}\ .
\end{equation}

The diagonal elements of the $N\times N$ matrix $\Phi^i$ are
associated with transverse space coordinates of the N Dp--branes
when they are separated.

The term $S_{WZ}$  in \p{1} is a Wess--Zumino term which describes
the coupling of the NDp--brane system to lower as well as to
higher--rank RR fields, the coupling to the latter is possible
because of the non--commutativity of the adjoint scalar fields
$\Phi^i$ \cite{Myers:1999ps,Taylor:2000pr}.

Finally, it should be noted that all the background fields in the
non--Abelian action \p{1} are assumed to be the same functionals
of the adjoint scalars $\Phi^i$ as they are in the Abelian theory
of a single Dp--brane \cite{douglas,hull}. Further details on the
structure of the action
\p{1} the reader may find in \cite{Myers:1999ps}.

Leaving apart a topical problem of the (symmetrized) trace of
the non--Abelian Dirac--Born--Infeld theory
\cite{Tseytlin:1997cs}, one can also be unsatisfied that the
action \p{1} is in the static gauge and, hence, is not worldvolume
reparametrization invariant. This action is also not invariant
under target space diffeomorphisms (except for the case of the
space--filling branes, i.e. pure non--Abelian DBI theories, where
there is no scalar fields), though it is invariant under the gauge
transformations of the background RR fields, as has been recently
proved in
\cite{ciocarlie}.

The lack of worldvolume and target space diffeomorphism invariance
may be a reason why there is a problem with supersymmetrizing the
multiple brane actions (in target space and on the worldvolume
(making them $\kappa$--symmetric)) and thus with introducing
worldvolume fermions into the construction. By now
supersymmetrization and fermion couplings have been considered
only for D0--branes in the context of Matrix theory
\cite{Taylor:2000pr} and for non--Abelian DBI theories
\cite{Ketov:2000fv,Bergshoeff:2001kt,Bergshoeff:2001dc}. In
\cite{Bergshoeff:2001kt} a possibility of constructing a
supersymmetric space--filling NDp--brane action with a
non--Abelian $\kappa$--symmetry has been considered up to $F^2$
terms in the variation, but then the conclusion has been made
\cite{Bergshoeff:2001dc} that such a construction would fail
beyond the quadratic approximation.

In this note we will address ourselves to studying the above
mentioned (super)symmetry problems of N coincident Dp--branes.

A natural way would be to start from the sum of N
reparametrization invariant and $\kappa$--symmetric actions for N
separated Dp--branes and analyze how their symmetries get modified
because of the contribution of the non--Abelian fields when the N
D-branes coincide\footnote{I am thankful to Arkady Tseytlin for
the discussion of this point}, but this program seems to be rather
complicated and ambitious, as has been also discussed in
\cite{Bergshoeff:2001kt}.

So to approach somehow these problems we suggest to look at the system
of N coincident Dp--branes from a somewhat different
point of view. Namely, let us consider them as a
qualitatively new
{\it single} brane configuration which is created when N
Dp--branes stack up together. We shall call it the NDp--brane. The
trace part
\begin{equation}\label{trace}
x^i(\sigma)={1\over{N}}{\rm Tr}\Phi^i
\end{equation}
 of the $U(N)$ adjoint scalars $\Phi^i$
(which is the center of mass of the N branes) is then naturally
identified with the transverse coordinates of this single brane
object in the target space--time. Together with the worldvolume
coordinates $\sigma^a$ they may be regarded as NDp--brane
coordinates $x^M(\sigma)=(\sigma^a, x^i(\sigma))$
$(m=0,1,\cdots, D)$ in a D--dimensional target space in the static
gauge. We can now leave the static gauge by making the theory
worldvolume diffeomorphism invariant introducing $p+1$ coordinates
$x^a(\sigma)$:
\begin{equation}\label{3}
x^M(\sigma)=(x^a(\sigma), x^i(\sigma)),
\quad (M=0,1,\cdots, D)\,.
\end{equation}
If we admit this point of view then the $U(N)$ vector fields
$A_a(\sigma)$ and the `traceless' scalars
\begin{equation}\label{4}
\varphi^i(\sigma)\equiv \Phi^i(\sigma)-x^i(\sigma) {\bf I}, \qquad \in
\qquad SU(N),
\end{equation}
which take values in $SU(N)$, should
be considered as pure worldvolume vector and scalar fields living
on the NDp--brane.

Thus, to make the action \p{1} worldvolume reparametrization
invariant we should rewrite it in a form where the center of mass
coordinates $x^i$ \p{trace} are separated from the $SU(N)$ scalar
fields $\varphi^i(\sigma)$ \p{4}.

Having written the NDp--brane action in a reparametrization
invariant form and having introduced the coordinates $x^M$ of the
target space, one may think of whether this construction can be
generalized to be invariant under target space supersymmetry (by
extending the target space to a superspace with Grassmann spinor
coordinates) and under worldvolume ({\it Abelian}) fermionic
$\kappa$--symmetry which transforms the target superspace
coordinates and ''matter'' fields on the brane\footnote{Remember that in the case of
the single superbranes the anticommutator of $\kappa$--symmetry
always generates worldvolume diffeomorphisms.}.

Note that in contrast to \cite{Bergshoeff:2001kt} we {\it a
priori} assume the NDp--brane action to be invariant under only
one (Abelian) $\kappa$--symmetry, since $N-1$ $\kappa$--symmetries
of the initially separated N super--Dp--branes are regarded to
have been gauge fixed.

The NDp--brane action \p{1} has a rather complicated form so to
simplify further analysis we shall restrict ourselves to a
particular class of N coincident Dp--branes of codimension one,
i.e. to the Dp--branes whose worldvolume has one dimension less
than the dimension of the target space ($p+1=D-1$).

In this case we have only one $U(N)$ adjoint scalar, hence all
commutators of $\Phi(\sigma)$ in \p{2} and \p{1} vanish and the
Wess--Zumino term does not have the contribution of couplings to
higher--rank RR fields thus reducing to the non--Abelian
generalization of the standard WZ term of a single Dp--brane
\cite{dbrane}, as in the case of the N coincident space filling D-branes.

In Section 2 we shall write a worldvolume reparametrization
invariant action for a codimension one NDp--brane and discuss its
symmetry properties. In Section 3 we shall simplify things even
further by considering a system of N coincident D0--branes in a
D=2 target space. We shall extend this system to an ND0--brane
propagating in a type IIA D=2 target superspace. The worldline
reparametrization invariant action describing the dynamics of this
system is shown to possess a target space supersymmetry, worldline
$\kappa$--symmetry and a number of linearly and non--linearly realized
rigid worldvolume supersymmetries. We find that the type IIA $D=2$
superalgebra generating the target space supersymmetry
transformations of the ND0--brane system acquires a non--trivial
``central extension" due to a nonlinear  contribution of the
$U(N)$ adjoint scalar fields. To the best of our knowledge this is
the first example of a target--space supersymmetric and
$\kappa$--invariant system of N coincident D--branes. In
Conclusion we comment on the possibility of extending these
results to higher--dimensional NDp--branes.

\section{The codimension one NDp--branes}

As we have discussed in the Introduction, in the case of
codimension one Dp--branes the action \p{1} reduces to
\begin{equation}\label{5}
{S}=-T_p \int d^{p+1}\sigma\,{\rm Tr}e^{-\phi}\sqrt{-det\left(
P[G_{ab}+B_{ab}]+ F_{ab}\right)}+ T_p\int {\rm Tr}P\left[ \sum
C^{(n+1)}\,e^B\right]e^{F}\ ,
\end{equation}
where $C^{(n+1)}$ are (n+1)--form RR potentials with
$n=p,p-2,p-4,\cdots$.

We now separate the transverse (center--of--mass) coordinate
$x^\bot(\sigma)$ from the $U(N)$ adjoint scalars and restore the worldvolume
reparametrization invariance of \p{5} by introducing p+1
coordinates $x^a(\sigma)$ as in \p{trace}--\p{4}. The action \p{5}
takes the form
\begin{eqnarray}\label{6}
{S}&=&T_p\int {\rm
Tr}P\left[\sum C^{(n+1)}\,e^B\right]e^{F}-T_p \int
d^{p+1}\sigma\,{\rm Tr}\,e^{-\phi}\\
 &&\times \sqrt{-det\left(
\partial_ax^M\partial_bx^NE_{MN}+ F_{ab}+
\partial_ax^ME_{M\bot}D_b\varphi+D_a\varphi
E_{\bot N}\partial_bx^N+ D_a\varphi
G_{\bot\bot}D_b\varphi\right)}.\nonumber
\end{eqnarray}
where $\bot$ stands for the index of the target space direction,
transverse to the brane worldvolume. Recall that in the case under
consideration we have a single $SU(N)$ adjoint scalar
$\varphi(\sigma)$.

Looking at the form of the action \p{6} we see that though we have
made it worldvolume reparametrization invariant the action has
remained non--invariant under target space diffeomorphisms. One
might expect this from the beginning, since, as we have already
mentioned, the background fields and hence the action for N
coincident Dp--branes depends on the U(N) adjoint scalars
$\Phi^i=x^i\,{\bf I}+\varphi^i$ and not separately on $x^M$ and
$\varphi^i$. In particular, this means that in the flat background
the NDp--brane action does not possess target--space Lorentz
invariance even in its center--of--mass part:
\begin{equation}\label{8}
S=- T_p \int d^{p+1}\sigma\,{\rm Tr}
\sqrt{-det\left(
\partial_ax^M\partial_bx^N\eta_{MN}+ F_{ab}+ D_a\varphi D_b\varphi+
2\partial_{(a}x^\bot D_{b)}\varphi\right)},
\end{equation}
where we have taken the background metric to be flat
$\eta_{MN}={\rm diag}(-,+,\cdots,+)$ and put all other background
fields to zero.

If we forget about the space--time origin of the SU(N) adjoint
fields $\varphi^i(\sigma)$ and (in accordance with our single
NDp--brane ideology) will consider them as  non--Abelian pure
worldvolume scalars, which does not transform under target space
Lorentz transformations but under an internal group SO(D-p-1),
then the only term which spoils the target space Lorentz
invariance of the NDp--brane action is the last term in \p{8}. Its
presence implies that the dynamics of the center of mass of the
system is directly affected by internal non--Abelian fluctuations
which is rather unusual.

If we drop this term we shall get a worldvolume diffeomorphism and
target space Lorentz invariant brane system with worldvolume U(N)
gauge fields
$A_a(\xi)$ and an
$SU(N)$ adjoint scalar field $\varphi$ propagating on the brane
\begin{equation}\label{9}
S=- T_p \int d^{p+1}\sigma\,{\rm Tr}
\sqrt{-det\left(
\partial_ax^M\partial_bx^N\eta_{MN}+ F_{ab}+ D_a\varphi D_b\varphi
\right)}.
\end{equation}

One may wonder whether these relativistic extended objects carrying
a non--Abelian matter on their worldvolumes may
find a place for themselves in String Theory. We shall discuss an
example of such systems in Section 4.

Turning back to the NDp--brane action \p{8}, we can notice that,
though being not Lorentz invariant, it is invariant under target
space rigid translations
\begin{equation}\label{10}
x^M~~\rightarrow~~x^M+a^M\,.
\end{equation}
This allows us to assume that by introducing fermionic degrees of
freedom the action \p{8} can be generalized to be invariant under
rigid space--time supersymmetry transformations whose
anticommutator closes on space--time translations. In addition,
this model, being generalized in a proper way, can also possess a
local fermionic $\kappa$--symmetry, whose anticommutator produces
worldvolume diffeomorphisms.

In the next Section, using a simple example of N D0--branes in D=2
space--time, we will demonstrate that such a generalization is
indeed possible.

\section{Supersymmetric and $\kappa$--symmetric ND0--brane model
in type IIA D=2 superspace}

\subsection{Bosonic ND0--brane system}
In the case of the system of N coincident D0--branes in D=2
space--time parametrized by two coordinates $x^M$ $(M=0,1)$ the
action \p{8} takes the form
\begin{equation}\label{11}
S=- m \int d\tau\,{\rm Tr}
\sqrt{-\left[
\dot x^M\dot x^N\eta_{MN}+(\dot\varphi)^2 +
2\dot x^1 \dot\varphi\right]},
\end{equation}
where $m$ is the mass of the single D0--brane (the mass of the N
D0--branes being, naturally, $Nm$),
$\tau$ is the worldline time parameter and  `dot' denotes its
derivative. Note that there is no place for the field strength of
the gauge field
$A(\tau)$ on the one dimensional worldline. Note also that since
only
$\dot\varphi$ enters the action, there is no ambiguity in choosing
the U(N) trace. It is automatically symmetric.

By construction the action \p{11} is worldline reparametrization
invariant, thus there must be the first class constraint which
generates this symmetry. This constraint should be a
generalization of (and reduce to) the mass shell condition for a
massive relativistic particle of a mass $Nm$, when
$\varphi=0$
\begin{equation}\label{12}
p_Mp^M+(Nm)^2=0,
\end{equation}
where $p_M$ is the particle momentum canonical conjugate to
$x^M=(x^0,x^1)$.

Let us find the reparametrization constraint for the ND0--brane
system.

 From the action \p{11} we derive the canonical momenta associated
with $x^0$ and $\Phi=x^1 {\bf I}+\varphi$, respectively,
\begin{equation}\label{13}
p_0={{\delta S}\over{\delta \dot x^0}}=-m\dot x^0{\rm Tr}
\left[\sqrt{-\left(
\dot x^2+\dot\varphi^2 +
2\dot x^1 \dot\varphi\right)}\right]^{-1},
\end{equation}
\begin{equation}\label{14}
p_\Phi={{\delta S}\over{\delta \dot
\Phi}}=m\dot\Phi\left[\sqrt{
(\dot x^0)^2-\dot\Phi^2 }
\right]^{-1}\equiv m\dot\Phi\left[\sqrt{-\left(
\dot x^2+\dot\varphi^2 +
2\dot x^1 \dot\varphi\right)}
\right]^{-1} \quad {\rm (no ~~trace!)},
\end{equation}
where $\dot x^2\equiv \dot x^N\dot x^N\eta_{MN}$. The momentum
conjugate to the spatial coordinate $x^1$ is the trace of
$p_\Phi$
\begin{equation}\label{15}
p_1= {{\delta S}\over{\delta \dot x^1}}=m {\rm
Tr}\left\{\dot\Phi\left[\sqrt{-\left(
\dot x^2+\dot\varphi^2 +
2\dot x^1 \dot\varphi\right)}
\right]^{-1}\right\},
\end{equation}
and the momentum associated with $\varphi$ is
\begin{equation}\label{16}
p_\varphi=p_\Phi-{1\over N}p_1\,{\bf I}.
\end{equation}

Upon some algebra we find the following constraint on the momenta
\begin{equation}\label{17}
-(p_0)^2+\left(Tr\sqrt{p_\Phi^2+m^2\,{\bf I}}\right)^2=0.
\end{equation}
As one can immediately check, in view of \p{16}, the constraint
\p{17} reduces to \p{12} when the adjoint field
$\varphi(\tau)$ is zero.

For further supersymmetrization it is convenient to rewrite the
action \p{11} in the first order form
\begin{equation}\label{18}
S=\int d\tau {\rm Tr}\left\{{1\over N}p_M\dot
x^M+p_\varphi\dot\varphi-{{e(\tau)}\over
2N}\left[-(p_0)^2+\left(Tr\sqrt{p_\Phi^2+m^2\,{\bf
I}}\right)^2\right]\right\},
\end{equation}
where $e(\tau)$ is the Lagrange multiplier ensuring the
constraint \p{17}.

We are now in a position to add to the system fermionic modes and
lift it to the type IIA $D=2$ superspace. To this end let us first
remind the structure and the symmetries of the action for a single
super--D0--brane, i.e. a massive type IIA  superparticle.

\subsection{The massive type IIA  superparticle}

The description of various aspects of the dynamics,
group--theoretical and geometrical properties of the massive
superparticle in type IIA $D=2$ superspace the reader may find e.g.
in \cite{achu,Pasti:2000zs}. Similar to the $D=10$ case the type
IIA $D=2$ superspace is parametrized, in addition to the bosonic
coordinates
$x^M$, by a two--component (real) Majorana spinor, or two Majorana--Weyl
coordinates of different chirality
\begin{equation}\label{19}
\theta^\alpha=(\theta^1,\theta^2).
\end{equation}

In the first order formalism the type IIA $D=2$ superparticle
action has the following form
\begin{equation}\label{20}
S=\int d\tau
\left\{p_M(\dot x^M+i\theta\gamma^M\dot\theta)-{{e(\tau)}\over
2}(p_Mp^M+m^2)+im\,\theta\gamma^2\dot\theta\right\},
\end{equation}
where the last term in \p{20} is the Wess--Zumino or the
Chern--Simons term, and $\gamma^M_{\alpha\beta}$ (M=0,1) and
$(\gamma^2)_{\alpha\beta}$ are $D=2$ Dirac
matrices in the Majorana representation:
\begin{equation}\label{gamma}
\gamma^{0}_{ \alpha \beta}= \left(
\begin{array}{cc}
1  &  0\\
0 &  1
\end{array}
\right)\,, \quad \gamma^{1}_{ \alpha \beta}=
\left(
\begin{array}{cc}
0  &  1\\
1 &  0
\end{array}
\right)\, , \quad \gamma^2_{ \alpha \beta}=\left(
\begin{array}{cc}
-1   &  0\\
0   &   1
\end{array}
\right) \,.
\end{equation}
The charge conjugation matrix $C$ for raising and lowering spinor
indices is
\begin{equation}\label{C}
 C_{ \alpha \beta}=C^{ \alpha \beta}=\left(
\begin{array}{cc}
0  &  1\\
-1 &  0
\end{array}
\right)\,.
\end{equation}

The action \p{20} is invariant under the $D=2$ Lorentz rotations,
Poincare translations, global supersymmetry
\begin{equation}\label{21}
\delta_\epsilon\theta^\alpha=\epsilon^\alpha, \qquad \delta_\epsilon
x^M=-i\epsilon\gamma^M\theta,\qquad \delta_\epsilon
P_M=0=\delta_\epsilon e
\end{equation}
and local fermionic $\kappa$--symmetry
\begin{equation}\label{22}
\delta_\kappa\theta=(p_M\gamma^M+m\gamma^2)\kappa(\tau), \quad
\delta_\kappa
x^M=i\delta\theta\gamma^M\theta,\quad \delta_\kappa
e=4i\kappa^\alpha\dot\theta_\alpha, \quad \delta_\kappa p_M=0,
\end{equation}
where $\kappa^\alpha(\tau)$ is a Grassmann--odd Majorana spinor
parameter of the $\kappa$--symmetry.

Note that due to the Dirac matrix algebra the square of
$(p_M\gamma^M+m\gamma^2)$ in the
$\kappa$--variation of $\theta$ is proportional to the mass shell
condition
\begin{equation}\label{23}
(p_M\gamma^M+m\gamma^2)(p_N\gamma^N+m\gamma^2)=(p_Mp^M+m^2)\,C\,.
\end{equation}
This means that only half of $\kappa^\alpha(\tau)$ effectively
contributes to the $\kappa$--variations \p{22}, and that
$\kappa^\alpha$ can eliminate only one of the components of
$\theta^\alpha$ by imposing, say a covariant gauge $\theta^1=0$.

In this gauge the action takes the form
\begin{equation}\label{24}
S=\int d\tau
\left\{-p^0(\dot x^0+i\theta^1\dot\theta^1)+p_1\dot x^1-{{e(\tau)}\over
2}(p_Mp^M+m^2)+ im\,\theta^2\dot\theta^2\right\}.
\end{equation}
Finally rescaling $\theta^2$ as
\begin{equation}\label{25}
\chi=\sqrt{(p^0-m)}\,\theta^2
\end{equation}
we arrive at the action with the free fermion kinetic term
\begin{equation}\label{26}
S=\int d\tau
\left\{-p^0\dot x^0+p_1\dot x^1-{{e(\tau)}\over
2}(p_Mp^M+m^2)- i\chi\dot\chi\right\}.
\end{equation}
We have now all ingredients to supersymmetrize the action for the
N coincident D0--branes by analogy with the single superparticle
action.

\subsection{The supersymmetric and $\kappa$--invariant ND0--brane action}

We assume that when the $\kappa$--symmetry is gauge fixed the
supersymmetric ND0--brane action has the form analogous to
\p{26}, but with the bosonic part as in eq. \p{18} and the Abelian
fermion $\chi$ being extended to a U(N) adjoint fermion
\begin{equation}\label{27}
\Psi(\tau)=\chi(\tau)\, {\bf I}+\psi(\tau),
\end{equation}
where $\psi(\tau)$ is a traceless fermionic matrix.

 We thus write the $\kappa$--gauge fixed super--ND0--brane action
 in the following form
$$
S=\int d\tau {\rm Tr}\left\{{1\over N}p_M\dot
x^M+p_\varphi\dot\varphi-{{e(\tau)}\over
2N}\left[-(p_0)^2+\left(Tr\sqrt{p_\Phi^2+m^2\,{\bf
I}}\right)^2\right]-i\Psi\dot\Psi \right\},
$$
or
\begin{equation}\label{28}
S=\int d\tau {\rm Tr}\left\{{1\over N}p_M\dot
x^M+p_\varphi\dot\varphi-{{e(\tau)}\over
2N}\left[-(p_0)^2+\left(Tr\sqrt{p_\Phi^2+m^2\,{\bf
I}}\right)^2\right]-i\chi\dot\chi-i\psi\dot\psi \right\}.
\end{equation}
Note that the action \p{28} does not contain non--diagonal
fermionic terms of the form $\chi\dot\psi$ which vanish because
$\psi$ is traceless. If we in addition impose the static gauge
$x^0=\tau$ the action reduces to
\begin{equation}\label{static}
S=\int d\tau {\rm Tr}\left\{p_\Phi\dot\Phi-
\sqrt{p_\Phi^2+m^2\,{\bf I}}-i\chi\dot\chi-i\psi\dot\psi \right\},
\end{equation}
where $H=Tr\sqrt{p_\Phi^2+m^2\,{\bf I}}$ is the Hamiltonian of the
gauge fixed system.

We would now like to restore the manifest global supersymmetry
\p{21} of the action \p{28} by adding
$\theta^1$ and interpreting $\chi$ as a $\theta^2$--component of
the Grassmann spinor coordinate $\theta^\alpha$ \p{19}. Upon
having appropriately rescaled $\chi$ and by analogy with \p{20} we
can write the following action
\begin{eqnarray}\label{29}
S=\int d\tau {\rm Tr}\left\{{1\over N}p_M(\dot
x^M+i\theta\gamma^M\dot\theta)+p_\varphi\dot\varphi-{{e(\tau)}\over
2N}\left[-(p_0)^2+\left(Tr\sqrt{p_\Phi^2+m^2\,{\bf
I}}\right)^2\right]\right.&&\nn  \left.+i{m}\,
\theta\gamma^2\dot\theta-i\psi\dot\psi \right\}.&&
\end{eqnarray}
The action \p{29} is invariant under the global supersymmetry
transformations \p{21} but it is not yet $\kappa$--symmetric. The
reason is that the mass shell condition \p{17} of the ND0--brane
system differs from the mass shell condition \p{23} of the single
superparticle. Let us rewrite \p{17} in the following form
\begin{equation}\label{30}
p_Mp^M+\left[\left(Tr\sqrt{p_\Phi^2+m^2\,{\bf
I}}\right)^2-(p_1)^2\right]=0.
\end{equation}
Comparing \p{30} with \p{23} we see that the term of \p{30} in the
square brackets plays the role of an ``effective mass" of the
ND0--brane system
\begin{equation}\label{31}
M(p_\varphi,p_1)\equiv \left[\left(Tr\sqrt{p_\Phi^2+m^2\,{\bf
I}}\right)^2-(p_1)^2\right]^{1/2}.
\end{equation}
Thus to restore $\kappa$--symmetry we should replace $m$ with ${M\over N}$
in the $\kappa$--symmetry variation rules \p{22} and in the term
$i{m}\theta\gamma^M\dot\theta$ of the action \p{29}.

The final form of the supersymmetric and $\kappa$--invariant
ND0--brane action is
\begin{equation}
\label{32}
S=\int d\tau {\rm Tr}\left\{{1\over N}p_M(\dot
x^M+i\theta\gamma^M\dot\theta)+p_\varphi\dot\varphi-{{e(\tau)}\over
2N}\left[p_Mp^M+M^2(p_\varphi,p_1)\right]\right. 
 \left.+i{M(p_\varphi,p_1)\over N}
\theta\gamma^2\dot\theta-i\psi\dot\psi \right\}.
\end{equation}
The action \p{32} is invariant under the following
$\kappa$--symmetry variations
\begin{equation}\label{33}
\delta_\kappa\theta=\left(p_M\gamma^M+M(p_\varphi,p_1)\gamma^2\right)\,\kappa(\tau),
\quad \delta_\kappa x^M=i\delta\theta\gamma^M\theta\,
+\,i\delta\theta\gamma^2\theta\,{{\delta
M(p_\varphi,p_1)}\over{\delta p_1}}\,\delta^M_1\,,
\end{equation}
$$
\delta_\kappa e=4i\kappa^\alpha\dot\theta_\alpha,
$$
\begin{equation}\label{kf}
\delta_\kappa\varphi=i\delta\theta\gamma^2\theta\,
\left[{{\delta M(p_\varphi,p_1)}\over{N\delta p_\varphi}}\right]_{traceless},
\end{equation}
where ${{\delta M(p_\varphi,p_1)}\over{\delta p_1}}$ and
${{\delta M(p_\varphi,p_1)}\over{\delta p_\varphi}}$ are
functions obtained by  varying  $M(p_\varphi,p_1)$ with
respect to $p_1$ and $p_\varphi$, respectively.
The $\psi$--fields are inert under the $\kappa$--symmetry.

We should note that since $M(p_\varphi,p_1)$, being a function
of $p_\varphi$ and $p_1$, is not constant the global supersymmetry
transformation of the spatial coordinate $x^1$ also gets modified (as under
$\kappa$--symmetry \p{33}) and
takes the form
\begin{equation}\label{x1}
\delta_\epsilon
x^1=-i\epsilon\gamma^1\theta-i\epsilon\gamma^2\theta\,{{\delta
M(p_\varphi,p_1)}\over{\delta p_1}},
\end{equation}

 This modified supersymmetry variation is
 a price for the model to be non--invariant under
$D=2$ Lorentz transformations. The Lorentz invariance of the first
order formulation is broken by an explicit dependence of $M$ on
$p_1$. In the next Section we shall consider a Lorentz--covariant counterpart
of this system with the standard
 supersymmetry transformations  of the type IIA superspace.

Because of the functional dependence of $M(p_\varphi,p_1)$ also
the $SU(N)$ adjoint scalar $\varphi$ transforms nontrivially under
the space--time supersymmetry, namely,
\begin{equation}\label{varphi}
\delta_\epsilon\varphi=-i\epsilon\gamma^2\theta\,
\left[{{\delta M(p_\varphi,p_1)}\over{N\delta p_\varphi}}\right]_{traceless}.
\end{equation}

The modified global supersymmetry transformations \p{x1},
\p{varphi} of $x^1$ and $\varphi$ close on generalized global bosonic
`translations' of $x^1$ and $\varphi$ under which the action
\p{32} is also invariant:
\begin{equation}\label{b}
\delta x^1=a^1+b\,{{\delta
M(p_\varphi,p_1)}\over{\delta p_1}}, \quad
\delta\varphi=A+b\,\left[{{\delta M(p_\varphi,p_1)}
\over{N\delta p_\varphi}}\right]_{traceless},
\end{equation}
where $a^1$ and $A$ are parameters of the standard `non--Abelian'
translations and $b$ is the parameter of the additional global
bosonic symmetry.

It is instructive to present the form of the supersymmetry algebra
of the transformations \p{x1}--\p{b} generated by the  Poisson
brackets of the Noether supercharges derived from the action
\p{32}
\begin{equation}\label{susya}
\{Q_\alpha,Q_\beta\}=2ip_M\gamma^M_{\alpha\beta}
+2iM(p_\varphi,p_1)\,\gamma^2_{\alpha\beta}\,,
\end{equation}
where
\begin{equation}\label{Q}
Q_\alpha=\pi_\alpha+ip_M(\gamma^M\theta)_\alpha +
iM(p_\varphi,p_1)\,(\gamma^2\theta)_\alpha\,,
\end{equation}
and $\pi_\alpha$ is the momentum conjugate to $\theta^\alpha$.

We observe that the superalgebra \p{susya} has the ``central
charge" term proportional to the effective mass
$M(p_\varphi,p_1)$ which arises because the $U(N)$ adjoint
scalars nontrivially transform
under supersymmetry. Such a ``central extension" of the
superalgebra is akin to that of the superalgebras associated with
the single superbranes having the Wess--Zumino terms in their
actions \cite{central} and gauge fields on their worldvolumes
which vary under target--space supersymmetry transformations
\cite{Sorokin:1997ps}.

When the adjoint scalar field
$\varphi(\tau)$ is switched off,
$p_\varphi=0$, $M(0,p_1)=Nm$ and the superalgebra
\p{susya} reduces to the standard type IIA superalgebra with the
central charge $Nm$ associated with the massive superparticle of
Subsection 3.2.

Finally the ND0--brane action \p{32} is invariant under the
following rigid worldvolume linear supersymmetry transformations
of
$\varphi$ and
$\psi$ with a U(N) adjoint fermionic parameter $\alpha$
\begin{equation}\label{abel}
\delta\psi=[\alpha p_\varphi]_{traceless},
\quad \delta \varphi=-2i[\psi\alpha]_{traceless}.
\end{equation}
If we assume that in eq. \p{32} the trace is symmetrized then the
action is also invariant under non--linearly realized non--Abelian
worldline supersymmetry transformations of $\psi$
\begin{equation}\label{nonl}
\delta\psi=\left[\beta({\bf I}+if(\tau)\psi\dot\psi)\right]_{traceless},
\end{equation}
where $\beta$ is a constant $SU(N)$ adjoint fermionic parameter
and $f(\tau)$ is an arbitrary adjoint scalar function. Under
\p{nonl} the adjoint fermion transforms as a Goldstone
field, thus the non--Abelian supersymmetry \p{nonl} is
spontaneously broken. These non--Abelian supersymmetries may be
regarded as a relic of the  space--time supersymmetry of (N-1)
D0--branes of the ND0--brane system whose $\kappa$--symmetries
have been gauge fixed.

We also notice that if in \p{32} we take the symmetrized trace the
action cannot acquire  higher order corrections in $\psi$ and
$\dot\psi$ since they vanish identically.

The abundance of the worldvolume rigid supersymmetries of the
ND0--brane action in $D=2$ is connected with its free fermion
structure. For NDp--branes with $p\geq 0$ and $D>2$ higher order
fermionic terms should appear in the action and worldvolume
supersymmetries should be related to the space--time
supersymmetry.

To conclude this section let us note that the form of the action
\p{32} agrees with a supersymmetric action for N D0--branes in Matrix
Theory \cite{Taylor:2000pr} reduced to $D=2$.

\section{The Lorentz invariant super--ND0--brane}

We now present a Lorentz invariant counterpart of the ND0--brane
model considered in the previous section.

If in \p{11} we skip the term $\dot x\dot\varphi$ the model
becomes $D=2$ Lorentz invariant. The reparametrization constraint
takes the Lorentz--covariant form
\begin{equation}\label{34}
p_Mp^M+\left(Tr\sqrt{p_\varphi^2+m^2\,{\bf I}}\right)^2=0,
\end{equation}
and so does the effective mass $M$
\begin{equation}\label{35}
M(p_\varphi)=Tr\sqrt{p_\varphi^2+m^2\,{\bf I}}\,.
\end{equation}

The supersymmetric and $\kappa$ invariant action for the Lorentz
invariant ND0--brane is easily deduced from \p{32}
\begin{equation}\label{36}
S=\int d\tau {\rm Tr}\left\{{1\over N}p_M(\dot
x^M+i\theta\gamma^M\dot\theta)+p_\varphi\dot\varphi-{{e(\tau)}\over
2N}\left(p_Mp^M+M^2(p_\varphi)\right) +i{M(p_\varphi)\over N}\,
\theta\gamma^2\dot\theta-i\psi\dot\psi \right\}.
\end{equation}
The variation properties of the  fields with respect to the
symmetries of the action \p{36} are the same as  written in
equations \p{21}, \p{33}--\p{nonl}, except that now the
spatial coordinate $x^1$ transforms in the standard way
under the $\kappa$--symmetry, $D=2$ supersymmetry
and translations. This is because now
$M(p_\varphi)$ does not depend on $p_1$. The model is Lorentz
invariant and the type IIA supersymmetry algebra is standard when
acting on $x^M$ but it is extended by the ``central" charge
$M(p_\varphi)$ (as in \p{susya}) due to the presence of the worldvolume
adjoint scalar $\varphi(\tau)$, which transforms non--trivially
under the target space supersymmetry and $\kappa$--symmetry.

\section{Conclusion}

In this paper we have examined a particular class of N coincident
Dp--branes of codimension one. We have seen that for these branes
the structure of the worldvolume action significantly simplifies.
We have then made the action invariant under worldvolume
diffeomorphisms. Though this has not restored target space Lorentz
symmetry, the action became invariant under the space--time
translations. The invariance of the action under worldvolume
diffeomorphisms and target space translations has allowed us to
assume that the system of NDp--branes of codimension one can be
generalized to include fermions in a target space supersymmetric
and worldvolume
$\kappa$--symmetric fashion.

We have demonstrated that such a generalization is indeed possible
by constructing the supersymmetric and
$\kappa$--invariant action which describes N coincident D0--branes
in type IIA $D=2$ superspace.

We have noticed that by dropping some terms in the NDp--brane
action and assuming the $SU(n)$ adjoint scalars to be pure
worldvolume fields we can make the action Lorentz invariant, and
presented the supersymmetric and $\kappa$--invariant action for a
Lorentz covariant system of N D0--branes.

A natural generalization of the results of this paper would be
to try to construct corresponding supersymmetric and $\kappa$--invariant
actions for ND--brane systems of higher dimension and codimension.
The prescription of how to act is prompted by the example of the
ND0--brane system. One should write down the action for bosonic N
Dp--branes in a worldvolume diffeomorphism invariant form, find
dynamical constraints which generate the worldvolume
diffeomorphisms and then try to modify in an appropriate way the
action and the projector matrix in the $\kappa$--symmetry transformations
of the corresponding single super--Dp--brane.

An  important and interesting problem to think about is Lorentz
(non)invariance of the systems of N coincident Dp--branes.
The formalism of Lorentz harmonics might be useful in studying this problem
\cite{harm}.

\bigskip
\noindent {\bf Acknowledgements.} The author would like to thank
G.~Arutyunov,  I.~Bandos, E.~Ivanov, S.~Krivonos, I. Oda,
A.~Pashnev, P. Pasti, M. Tonin, A. Tseytlin and S. Zhukov for
interest to this work and illuminating discussions. This work was
partially supported by the European Commission TMR Programme
ERBFMPX-CT96-0045 to which the author is associated with the
University of Padua, by the Grant N 383 of the Ukrainian State
Fund for Fundamental Research and by the INTAS Research Project N
2000-254.

\end{document}